\newcommand{\be}{\begin{equation}}
\newcommand{\ee}{\end{equation}}

\documentclass[journal,10pt]{IEEEtran}

\usepackage{psfrag}
\ifCLASSINFOpdf
   \usepackage[pdftex]{graphicx}

\else
  
   \usepackage[dvips]{graphicx}

\fi
\usepackage[cmex10]{amsmath}
\usepackage{amsthm,epstopdf,nicefrac,pgfplots}

  \pgfplotsset{compat=newest} 
  \pgfplotsset{plot coordinates/math parser=false}

\usepackage{rotating}

\usepackage{mathtools}

\usepackage{gensymb,gnuplottex}

\usetikzlibrary{intersections,backgrounds}
\usepgfplotslibrary{fillbetween}

\renewcommand\appendix{\par
  \setcounter{section}{0}
  \setcounter{subsection}{0}
  \setcounter{figure}{0}
  \setcounter{table}{0}
  \renewcommand\thesection{Appendix \Alph{section}}
  \renewcommand\thefigure{\Alph{section}\arabic{figure}}
  \renewcommand\thetable{\Alph{section}\arabic{table}}
}

\usepackage{color}

\twocolumn

\usepackage[caption = false]{subfig}

  \newlength\fheight 
    \newlength\fwidth 
\usepackage{balance}
\usepackage{algorithm}

\usepackage{setspace}

\usepackage[bookmarks=false]{hyperref}
\newcommand\copyrighttext{%
  \footnotesize \textcopyright 2016 IEEE.  Personal use of this material is permitted. Permission from IEEE must be obtained for all other uses, in any current or future media, including reprinting/republishing this material for advertising or promotional purposes, creating new collective works, for resale or redistribution to servers or lists, or reuse of any copyrighted component of this work in other works. 
 DOI: \href{https://doi.org/10.1109/TIM.2016.2540865}{10.1109/TIM.2016.2540865}
 }
\newcommand\copyrightnotice{%
\begin{tikzpicture}[remember picture,overlay]
\node[anchor=south,yshift=10pt] at (current page.south) {\fbox{\parbox{\dimexpr\textwidth-\fboxsep-\fboxrule\relax}{\copyrighttext}}};
\end{tikzpicture}%
}
\begin{document}
%
\title{Measuring The Noise Cumulative Distribution Function Using Quantized Data}
%
%
%


\author{P.~Carbone,~\IEEEmembership{Fellow Member,~IEEE}\thanks{P. Carbone and Antonio Moschitta are with the University of Perugia - Engineering Department, via G. Duranti, 93 - 06125 Perugia Italy,}
and~J.~Schoukens,~\IEEEmembership{Fellow Member,~IEEE}\thanks{J. Schoukens is with the Vrije Universiteit Brussel, Department ELEC, Pleinlaan 2, B1050 Brussels, Belgium.}
and~I.~Koll\'ar,~\IEEEmembership{Fellow Member,~IEEE}\thanks{I. Koll\'ar is with the Budapest University of Technology and Economics, Department of Measurement and Information Systems, 1117 Budapest, Hungary.}
and~A.~Moschitta}

\maketitle
\copyrightnotice
\begin{abstract} This paper considers the problem of estimating the cumulative distribution function and probability density function 
of a random variable using data quantized by uniform and non--uniform quantizers.
A simple estimator is proposed based on the empirical 
distribution function that also takes the values of the quantizer transition levels into account. 
The properties of this estimator are discussed and analyzed at first by simulations. Then, 
by removing all assumptions that are difficult to apply, a new procedure is described that does not require neither 
the transition levels, nor
the input sequence used to source the quantizer to be known.  Experimental results obtained using a commercial $12$-bit data acquisition system show the applicability of this estimator to real-world type of problems.


\end{abstract}



%
\IEEEpeerreviewmaketitle


\section{Introduction}
Several of the
techniques used in 
electronic engineering require knowledge of the parameters
of the noise affecting systems and signals.
This is the case, for instance, when electronic systems 
need to extract information from signals acquired
by using data acquisition systems (DAQs) or analog-to-digital converters (ADCs). 
Accordingly, the noise cumulative distribution function (CDF)  and probability 
density  function (PDF) 
provide the user with enough information to better tune the used algorithms and estimators.

Other practical situations requiring knowledge of 
the input noise PDF occur when testing mixed-signal devices such as ADCs.
The characterization of the behavior of ADCs and DAQs done for verification purposes, includes measurement of the input noise basic properties, such as mean value or standard deviation \cite{IEEE1241}. Having the possibility to extend this knowledge by adding information on its CDF and PDF removes the need for making simplifying and unproven assumptions about the noise statistical behavior.

Also, in practice,
quantization is always affected by some additive noise contributions at the input. Noise may be  artificially added, as when dithering is performed \cite{AD9265},
or just be the effect of input-referred noise sources associated 
to the behavior of electronic devices. 
It is known that a small amount of additive noise added before quantization may linearize, on the average, 
the  stepwise input-output characteristic, but that a large amount of noise is needed for the linearization of quantizers 
with non-uniformly distributed transition levels \cite{Wagdy}\cite{CarboneNarduzziPetri}. 
Being able to characterize the input noise distribution is thus necessary if the user wants to have proper control over the whole acquisition chain, with or without the usage of dithering.

The problem of estimating the noise CDF and PDF based on quantized data is addressed in this paper.
Research results on this topic appear in \cite{Lee} where parametric identification of the input PDF based on quantized signals
was described in the context of image restoration. Similarly, in \cite{CarboneMoschittaPDF}, a parametric approach is taken to estimate 
the input PDF from quantized data. A solution to the same problem,
{\color{black} in the context of ADC testing}, is described in \cite{IEEE1241}.
Besides being central in measurement theory,
this problem is related to a class of {\color{black} control theory and system identification}
problems \cite{Chiuso}-\nocite{WangYinZhangZhao}\cite{104}, and is 
analyzed in the context of categorical data analysis  \cite{Qli}\cite{Agresti}. 

In this paper, we describe and characterize a simple {\color{black} nonparametric} estimator for the CDF and PDF of the noise at the input of a memoryless quantizer, not however required to have uniformly distributed transition levels. 
When using ADCs it is common that output codes are used for data processing and estimation purposes.
However, this is just one of the possible approaches.
In fact, if the quantizer threshold levels are known or measured,
the quantizer itself can be considered as a simple measuring 
instrument capable to assess the interval boundaries to which the input belongs. 
Thus, by processing data in the amplitude domain and by avoiding {\color{black} the usage of} quantizer codes, a possible increase in estimators' accuracy is obtained.  
To prove the validity of this approach, simulated data will be used, as well as results of a practical experiment based on equipment typically present in any laboratory environment.
Outcomes prove that the
proposed estimator is accurate and 
easily adoptable for practical purposes.

\section{A CDF Estimator}
\label{sec:esti}

\begin{figure}[t]
{\includegraphics[scale=0.25]{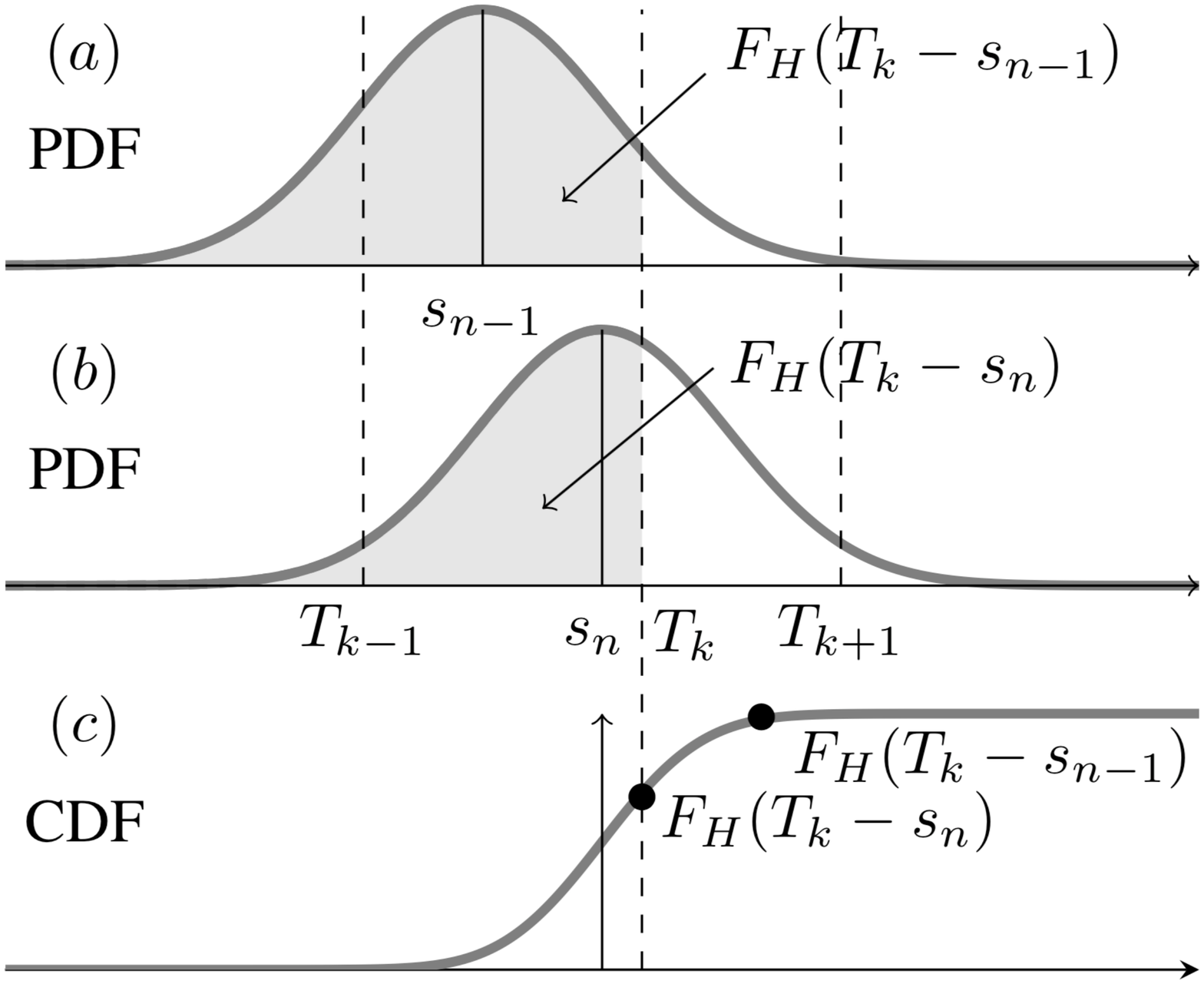}}
\caption{\color{black} 
Geometrical 
representation of the relationship between the PDF of the input signal $x(\cdot,\cdot)$ at times $n-1$ and $n$, and the CDF $F_H(\cdot)$ of
the additive noise, for a given value of the transition level $T_k$ (shaded areas), and assuming two different values of the deterministic sequence: $s_{n-1}$ (a) and $s_n$ (b). Consequently $F_H(\cdot)$ is evaluated at $x_{j-1}=T_k-s_{n-1}$ and $x_j=T_k-s_n$. Positions of neighboring transition levels $T_{k-1}$, and $T_{k+1}$ are also shown. Shaded areas representing theoretical probabilities are estimated simply by counting the relative number of occurrences of codes below $T_k$.
\label{figpdf}}
\end{figure}

Consider a quantizer having $K+1$ transition levels $T_k$ such that
it outputs $Q_k$, once the input value belongs to the interval
$[T_{k-1}, T_{k})$, where $k \in {\cal K}=\{k | k=1, \ldots, K\}$ represents the set of possible values for $k$ {\color{black} and $Q_1, \ldots, Q_K$ is an increasing sequence of  values}. Additionally assume that $R$ records of $N$ samples each 
are sourced to the quantizer, 
where each sequence 
can be written as
\be
	x(n,r) = s_n+\eta(n,r) \quad n=0, \ldots, N-1, \quad r=0, \ldots, R-1
\ee
with $n$ as the time index, $r$ as the record index,
$s_{n}$ as a deterministic 
sequence, and where $\eta(n,r)$ 
represents the $n$-th outcome of a noise sequence
in the $r$-th record, {\color{black} with independent outcomes and}
having PDF, $f_{H}(\eta)$. 
Additionally assume that the noise can be modeled as a stationary random process 
with properties that are independent from the input sequence $s_n$.
Accordingly, the noise properties do not depend on the quantizer level in an ADC.
Further assume that $x(n,r)$ is quantized by a 
quantizer within an {\color{black}ADC} or a {\color{black} DAQ}.
The sequence $y(n,r)$ of quantized values {\color{black} $Q_k$} is collected and processed by the experimenter to obtain an estimate $\hat{F}_H(\cdot)$ of the noise CDF $F_H(\cdot)$.
{\color{black} Then, the CDF $F_{X_n}(\cdot)$ of the input signal $x(\cdot,\cdot)$ at time $n$, 
is given by:
\begin{align}
\begin{split}
	F_{X_n}(T_k) & = P\left(
	x(n,r) \leq T_k\right) = P\left( s_n+\eta(n,r) \leq T_k\right) \\
	 & = F_H(T_k-s_n)
\end{split}
\end{align}
which shows the relationship between 
the noise CDF 
$F_H(\cdot)$ and $F_{X_n}(\cdot).$}
{\color{black} To highlight the behavior of  this relationship 
consider the PDFs and the CDF depicted in Fig.~\ref{figpdf}. Subfigures (a) and (b) show the behavior of the
PDFs of $x(\cdot,\cdot)$, assuming two different values $s_n$ and $s_{n-1}$ of the deterministic input sequence. 
Shaded areas in subfigures (a) and (b) correspond to the probability of
having experimental occurrences of $x(\cdot,\cdot)$ below the transition level $T_k$, that is $F_H(T_k-s_n)$ and $F_H(T_k-s_{n-1})$, respectively. Since these probabilities can easily be estimated by counting the number of times the ADC outputs a code lower than or equal to $Q_k$, an estimator of $F_H(\cdot)$ results in correspondence to two different values of its argument.
By taking into account the effect of input values $s_n$ and $s_{n-1}$, 
subfigure \ref{figpdf}(c) shows that $F_{X_{n-1}}(x)=F_H(T_k-s_{n-1})$ and 
$F_{X_n}(x)= F_H(T_k-s_n)$.
} 

To obtain an expression for the 
estimator of $F_H(T_k-s_n)$
observe that the difference 
$T_k-s_n$ may provide the same value for different combinations
of $n$ and $k$. Define $x_j$ as the ordered sequence of such values.
Thus, let us partition the set of all possible combinations $(n,k)$ 
in subsets ${\cal S}_{j}$  
such that if two couples $(n_1, k_1)$
and $(n_2,k_2)$ belong to ${\cal S}_{j}$, 
with $n_1 \neq n_2$ and/or $k_1 \neq k_2$,
then $T_{k_1}-s_{n_1}=T_{k_2}-s_{n_2}=x_j$. 
Define $L_j$ as the cardinality of ${\cal S}_j$ and 
 $L$ as the number of such subsets. Then $L=NK$ only if all subsets contain just one element, 
that is all differences $T_k-s_n$ provide unique values. 
Otherwise $L<NK$ and there is at least one subset ${\cal S}_j$ that
has cardinality larger than $1$. 
In both cases, $\cup_{j=1}^{L}{\cal S}_j$ contains all $NK$ 
possible combinations of $n$ and $k$.

An estimator of $F_H(x_j)$
is obtained by the empirical cumulative distribution function (ECDF) evaluated at $x_j$, that is 
\begin{align}
\begin{split}
	\hat{F}_H(x_j) = \frac{1}{RL_j}\sum_{
	\{ (n,k)|(n,k)\in {\cal S}_j \}}\sum_{r=0}^{R-1} &
	[ y(n,r) \leq Q_k ] \\
	&
	\quad j=1, \ldots, L
	\label{first}
	\end{split}
\end{align}
where $[A]$ is the indicator function of the event $A$, that is
$[A]=1$ if $A$ is true and $0$ otherwise.  
{\color{black} Thus, (\ref{first}) really defines a set  of ECDFs, each one calculated for a given value of $x_j$. 
{\color{black}The partitioning is useful, because all collected samples belonging to a set ${\cal S}_j$ provide information about the noise CDF value when its argument is $x_j$. Since $x_j$ can assume $L$ distinct values, the CDF is sampled at $L$ distinct points.}}
Observe also that 
the values $x_j$ in the argument of $\hat{F}_H(\cdot)$ may not be uniformly distributed over the 
input noise amplitude range and that the estimated CDF is independent of the actual values associated to $Q_k$.

To obtain an expression of the 
CDF estimator for any value of $x$, 
$\hat{F}_{H}(x_j)$ can be either 
interpolated between neighboring values using suitable interpolating functions, 
or a parametric model can be fitted to available data as shown in subsection \ref{subs:simulation}.
Finally, the CDF can be {\color{black}differentiated} either numerically or 
analytically to obtain
an estimate of the noise PDF.

\subsection{Estimator properties}
Estimator (\ref{first}) has
statistical properties that depend on whether {\color{black} the sequence} $x_j$ is assumed to be known or not.
\subsubsection{Known test parameters}
Knowledge of the sequence $x_j$ requires knowledge of both 
the transition level values $T_k$ and of the samples in the input sequence $s_n$.
If both are assumed to be known, the only source of variability is associated to sampling of the noise sequence. 
In this case, the estimator mean value and variance can be derived by observing that
$\hat{F}_H(x_j)$ is a binomial random variable. 
Thus, {\color{black} its mean value can  be calculated from (\ref{first})
by observing that 
\begin{align}
\begin{split}
	E\left( \left[ y(n,r)  \leq Q_k \right] \right)  & = 
  P\left( \left[ y(n,r) \leq Q_k \right] =1  \right)  \\
&  =	P\left( x(n,r) \leq T_k \right)
	 = F_{X_n}(T_k) \\
	 & = F_H(T_k-s_n)
	  =F_H(x_j)
	\label{meanfirst}
\end{split}
\end{align}
for some $j$ in $1, \ldots, L$ and regardless of $r$, so that
\begin{align}
\begin{split}
	E(\hat{F}_H(x_j)) = 
	\frac{1}{L_j}\sum_{
	\{ (n,k)|(n,k)\in {\cal S}_j \}}  F_H(x_j) & =
	F_H(x_j) \\ & \qquad j=1, \ldots, L
	\label{avg}
\end{split}
\end{align}
The variance of (\ref{first}) depends on the variance of $[y(n,r) \leq Q_k]$ given by
\begin{align}
\begin{split}
	\mbox{var}([y(n,r) \leq Q_k]) & = 
	P\left( \left[ y(n,r) \leq Q_k \right] =1 \right) \\
	&
	\times
	\left( 1
	-P\left( \left[ y(n,r) \leq Q_k \right] =1 \right)
	\right) \\ & = 	F_H(x_j)\left( 1- 	F_H(x_j) \right) \quad j=1, \ldots, L
\end{split}
\end{align}
so that
\begin{equation}
	\mbox{var}(\hat{F}_H(x_j)) = \frac{1}{RL_j} F_H(x_j) \left( 1-F_H(x_j) \right) \qquad j=1, \ldots, L
\label{varfirst}
\end{equation}
results because of the statistical independence hypothesis.
Expressions (\ref{meanfirst})-(\ref{varfirst}) show that the 
estimator is unbiased for selected values $x_j$ of its argument. Thus, it estimates correctly the unknown CDF by sampling it at $x=x_j$, $j=1, \ldots, L$. If $L$ is sufficiently large, the sampling process returns sufficient information for interpolating between neighboring samples to obtain an estimate of the CDF for any value of $x$. In the following, simple linear interpolation is adopted that is shown to provide acceptable results.
Many other published techniques can be found in the scientific literature \cite{Kurt}-\nocite{Masry}\cite{Parzen}.  
}
\subsubsection{Unknown test parameters}
When $x_j$ is unknown it must first be estimated.
This requires both the estimation of the DAQ transition levels and of 
the DAQ input samples in $s_n$. If unknown,  
the transition levels can be estimated using the procedures described in 
\cite{IEEE1241}, both under the hypothesis of DC and AC input signals.
Similarly, the input sequence $s_n$ can be determined
either by measuring it with a reference instrument at the DAQ input or by 
estimating it,
using the same quantized data at the DAQ output. 
\vskip0.2cm
\noindent
{\em Estimating $s_n$ using a swept DC or an AC signal}
\vskip0.05cm
If $s_n$ is first measured by a reference instrument, e.g., a digital multimeter (DMM) in the case of DC signals, it is affected by  measurement uncertainty, whose effect  
can be neglected if it is 
much smaller than the resolution of the considered DAQ.  
The DMM can be used 
for instance when $s_n$ is a sequence of swept DC values 
that are applied in sequence to the DAQ input for scanning the noise PDF.

Alternatively, an AC input can be applied, e.g., a sine wave.
In this case, the DMM can only be used for measuring general parameters such as the sine wave amplitude, but can not provide information on the single sample $s_n$. Thus, either a calibrated DAQ with a much smaller resolution is adopted as a reference instrument or, more practically, the sequence $s_n$ is estimated from the stream of DAQ output data.  This latter case is equivalent to considering the DAQ itself as the reference instrument used to obtain information about the input sequence.
A straightforward technique to be used for estimating 
$s_n$ in this case is the {\em sine-fit}, that is the least-squares approach.
Observe that if the input noise standard deviation is small compared to the quantization step, the sinefit is known to be a biased estimator \cite{Alegria}-\nocite{HandelLSE}\cite{CarboneSchoukens}. Thus, even if a large number of records is collected a biased noise CDF estimator will however result. 
Other estimators can be used to remove this bias at the expense of added complexity \cite{Kollar1}\cite{Kollar2}.

\vskip0.2cm
\noindent
{\em Dealing with limited knowledge about the needed information}
\vskip0.05cm
When the sequence $x_j$ is unknown either
because $T_k$, $s_n$ or both are unknown,
estimation, measurement, or quantization errors 
affect both the DAQ input and output sequences. 
When this is the case, the estimator can be reformulated as follows. 
Consider
\begin{align}
\begin{split}
\hat{x}_j & = \hat{T}_k-\hat{s}_n = x_j + \epsilon_x, \\
\end{split}
\label{EIV}
\end{align}
where $\epsilon_x$
models the uncertainty affecting knowledge of both $T_k$ and $s_n$, needed to obtain $x_j$.
Assume also
\begin{equation}
	\hat{F}_H(x_j) = F_H(x_j)+\epsilon_y
	\label{due}
\end{equation}
where $\epsilon_y$ represents the 
estimation error accounting 
for the effect of sampling {\color{black} over the available finite record of data}.
As shown in (\ref{avg}),  $\epsilon_y$ 
is a zero-mean random variable with variance
given by (\ref{varfirst}). 
Then, the updated noise estimator of $F(x_j)$ becomes 
$\hat{F}_H(\hat{x}_j)$, where $\hat{F}_H(\cdot)$ is calculated using
(\ref{first}) in which every occurrence of the unknown quantities $T_k$ and $s_n$ is replaced by the corresponding estimated or measured values.

\begin{figure}[t]
{\includegraphics[scale=0.45]{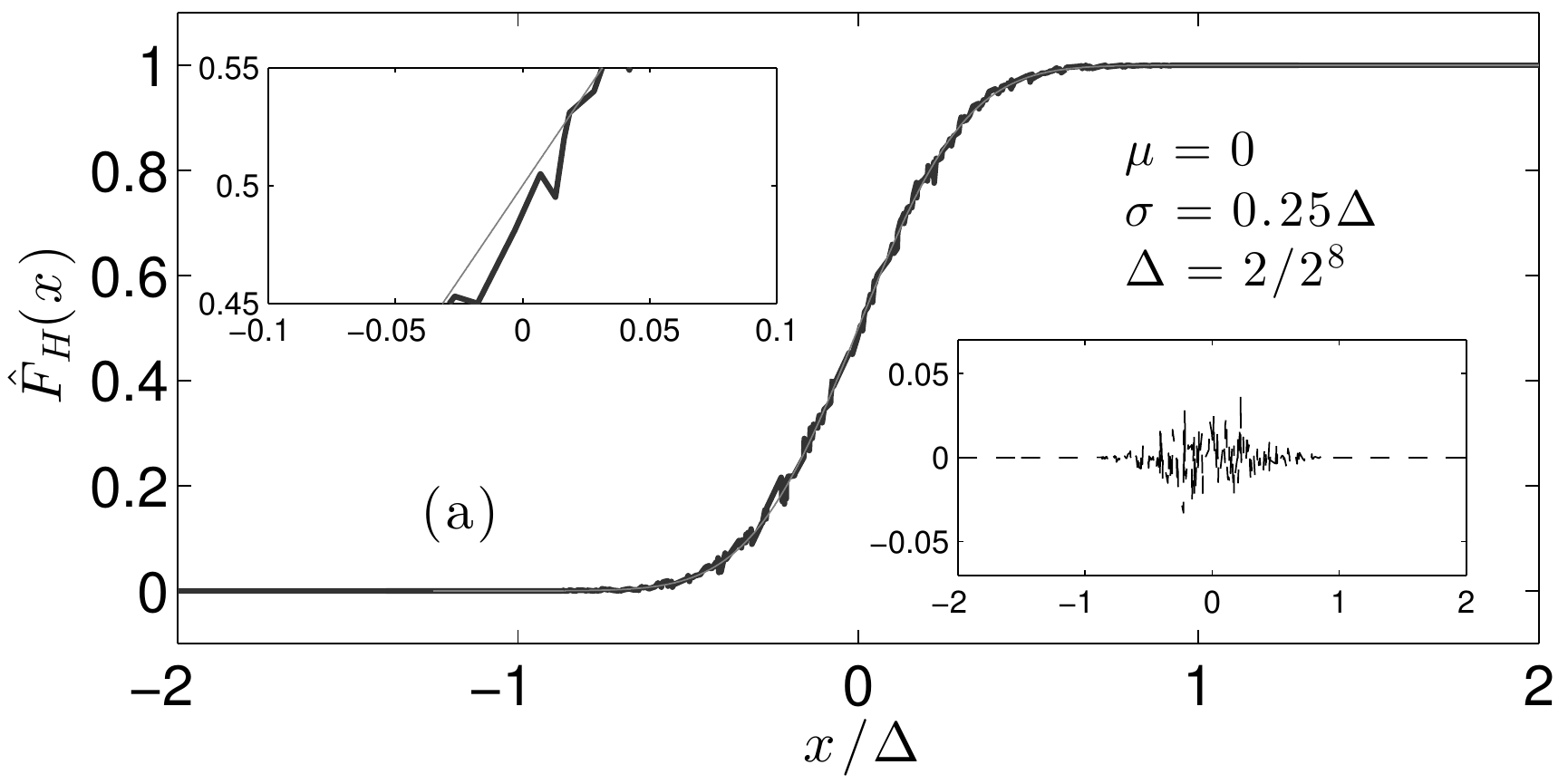}}
{\includegraphics[scale=0.45]{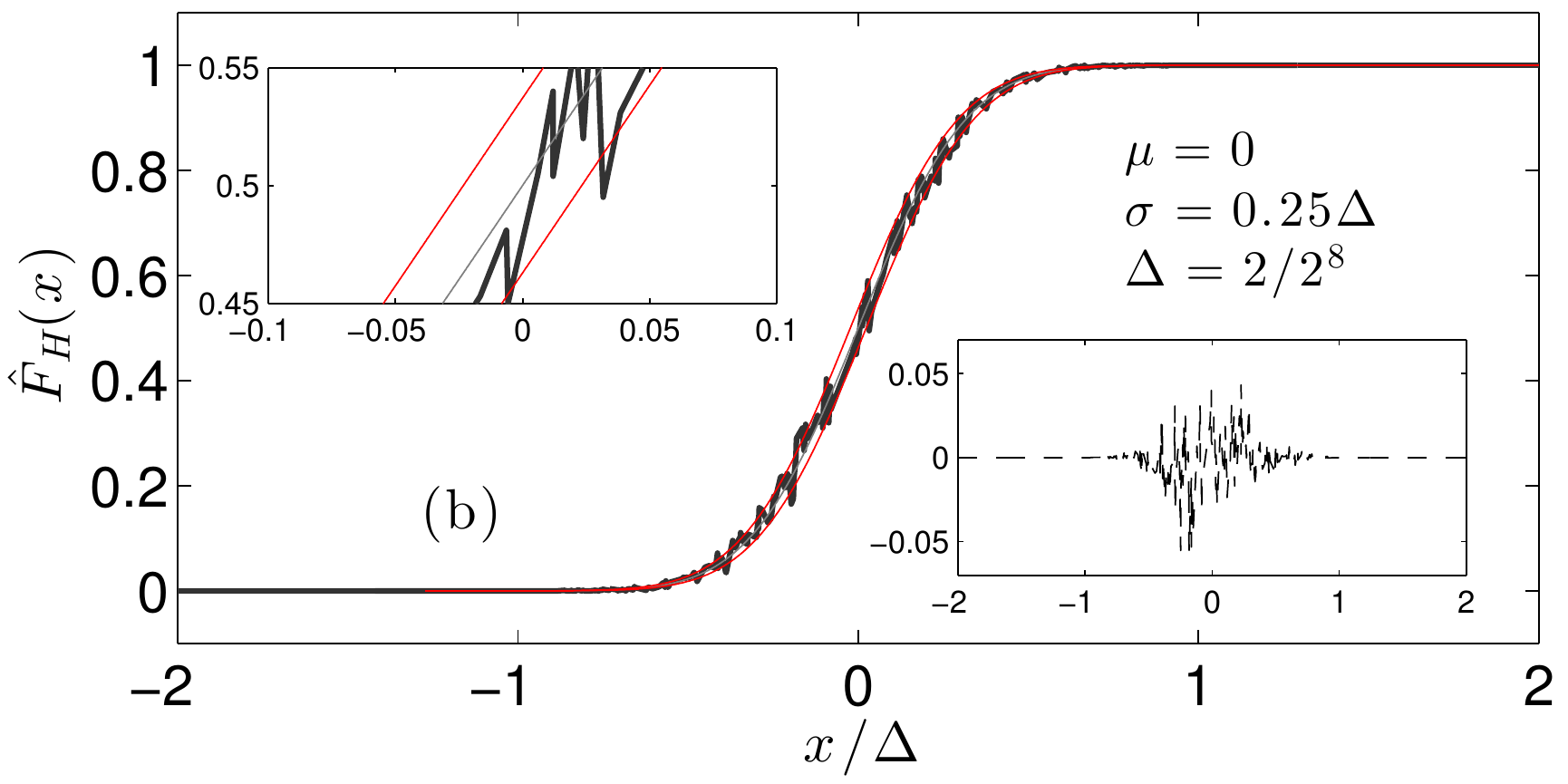}}
{\includegraphics[scale=0.45]{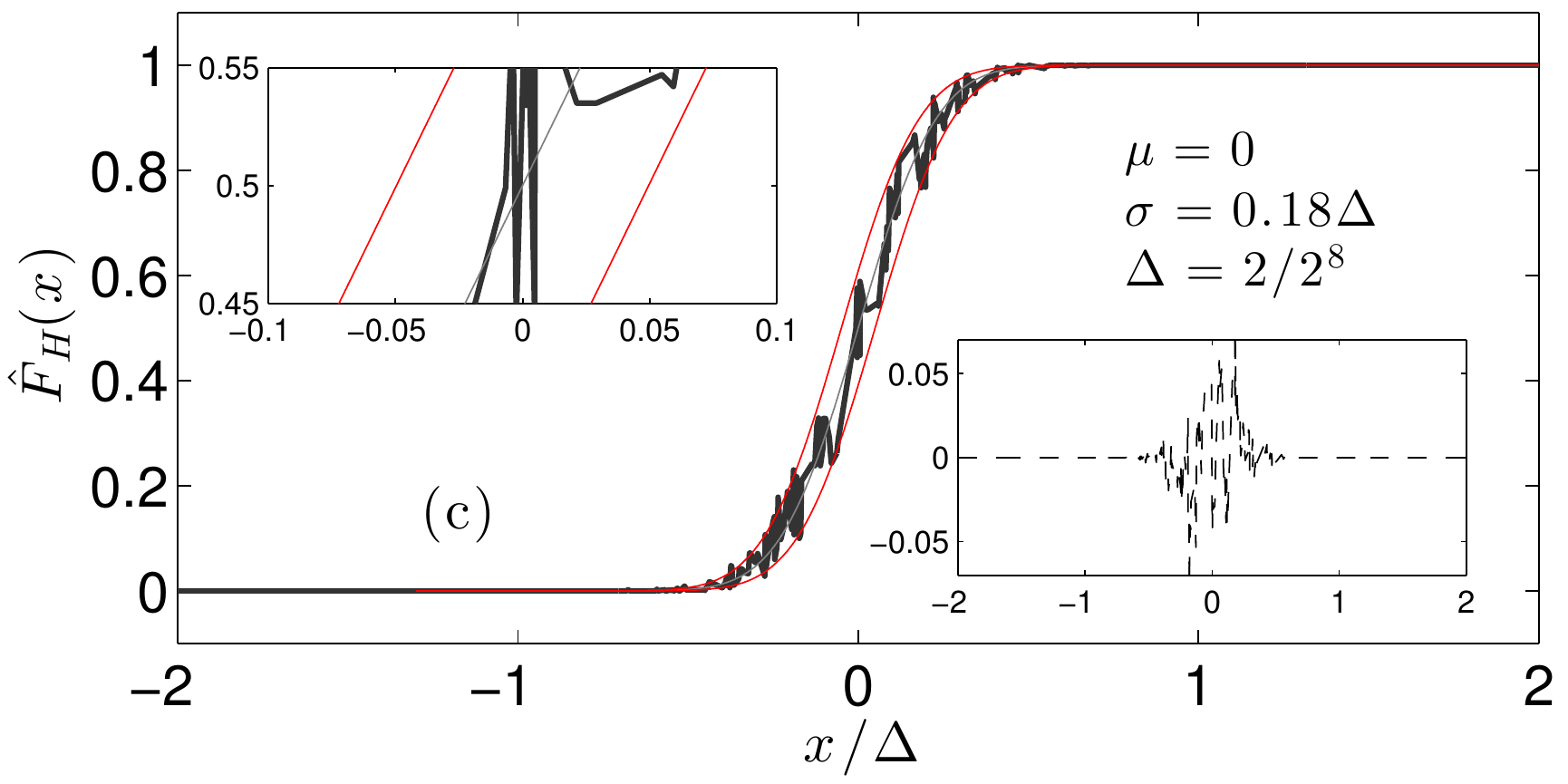}}
\caption{Simulation results: Estimation of the CDF of a zero-mean Gaussian random variable based on $R=1000$ records of data quantized by an $8$-bit uniform quantizer, when the input signal is a  
sine wave coherently sampled at $N=151$ different phase angles. 
Insets show a zoomed portion of the estimated (bold line) and true (thin line) CDF (upper left corner) and of the estimation error (lower right corner).
(a) Input sequence $s_n$ and transition levels $T_k$ assumed to be known; (b) Known transition levels and input sequence $s_n$ estimated by applying the sine fit procedure to the quantizer output data when 
the noise standard deviation is equal to $\sigma=0.25\Delta$, where $\Delta$ represents the nominal quantization step; (c) Same as (b) but with $\sigma=0.18\Delta$. Red lines in (b) and (c) represent approximate bounds for the estimation errors due to
the bias in estimating $s_n$ (see text).
\label{gau}}
\end{figure} 

\begin{figure}[t]
{\includegraphics[scale=0.45]{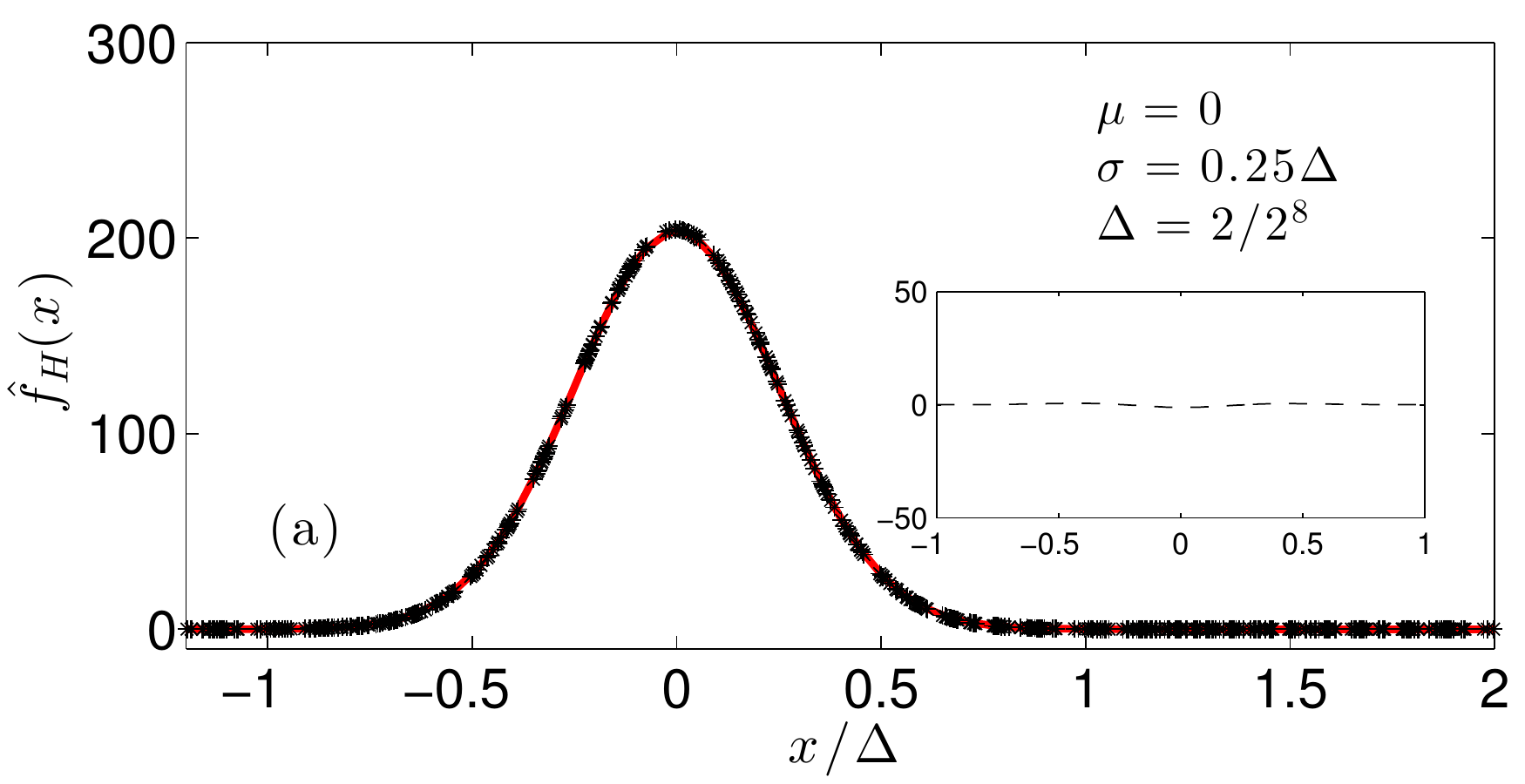}}
{\includegraphics[scale=0.45]{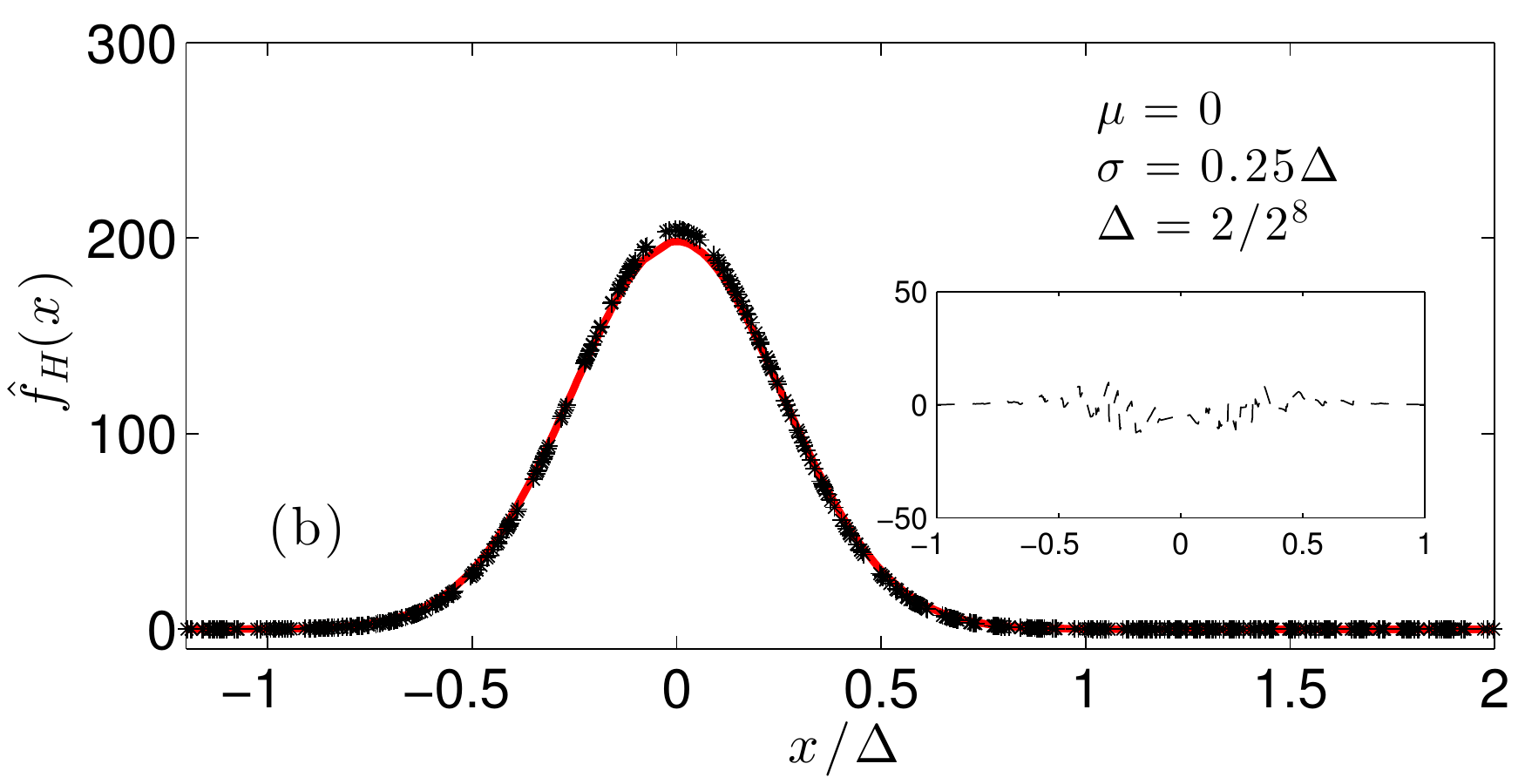}}
{\includegraphics[scale=0.45]{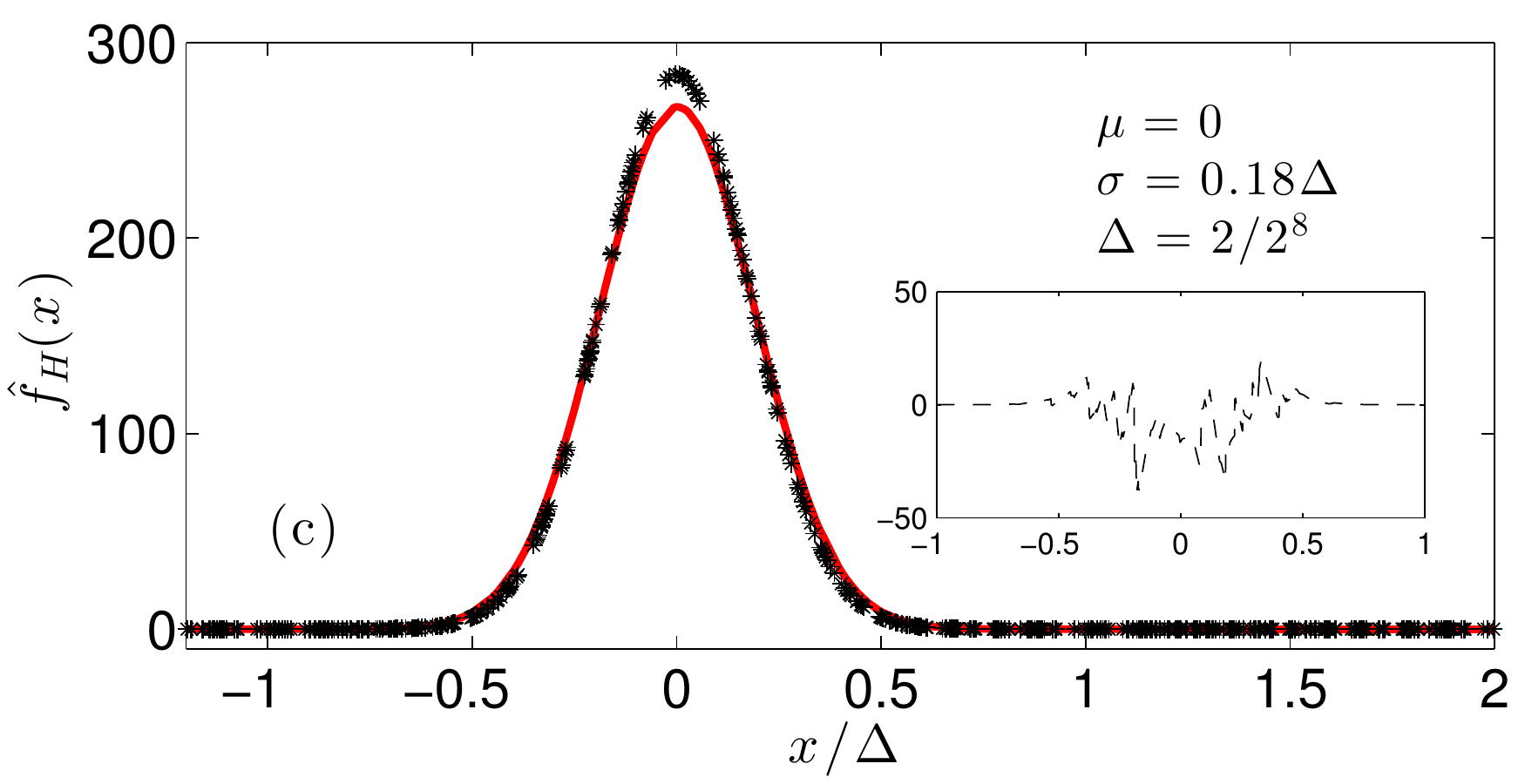}}
\caption{Simulation results: Estimation of the PDF of a zero-mean Gaussian random variable based on $R=1000$ records of data quantized by an $8$-bit uniform quantizer, when the input signal is a  
sine wave coherently sampled at $N=151$ different phase angles. 
Insets show a zoomed portion of the estimation error. 
Black stars represent the true PDF sampled at $s_n$. Red lines represent the estimated PDF.
(a) Input sequence $s_n$ and transition levels $T_k$ assumed to be known; (b) Known transition levels and input sequence $s_n$ estimated by applying the sine fit procedure to the quantizer output data when 
the noise standard deviation is equal to $\sigma=0.25\Delta$, where $\Delta$ represents the nominal quantization step; (c) Same as (b) but with $\sigma=0.18\Delta$. 
\label{montegau}}
 \end{figure}

Expressions in (\ref{EIV}) show that if the variance of $\epsilon_x$ and
of $\epsilon_y$ can be neglected and
$|\epsilon_x| < \Delta_{\epsilon}$, where
$\Delta_{\epsilon}$ represents a bound on the bias in the estimation of $x_j$,
then the estimated noise CDF is approximately bounded by $F(x_j-\Delta_{\epsilon})$ and $F(x_j+\Delta_{\epsilon})$. 
In the next section this property will {\color{black} further be} analyzed.

\section{Applying the estimator}
In this section we show how to apply the proposed estimator
both using simulations and experimental results.
Practical issues arise when (\ref{first}) is applied, as some 
of the needed information may not be available. 
Typically 
the values of $T_k$ and/or $s_n$ 
may not be known to the user or 
the construction of subsets ${\cal S}_j$ 
may not be straightforward,
owing to the approximate equivalence of the difference defining $x_j$. 
Details are given in the following subsections.

\subsection{Choice of $N$ and $R$}
The number of samples $N$ and of records $R$ have different consequences on the estimator performance.
$N$ determines the number of points at which the CDF is sampled.
Potentially $N$ different sampling point results. However 
the actual sampling point is $x_j=T_k-s_n$. Thus, different couples
$T_k, s_n$ can result in the same value $x_j$ as explained in Section~\ref{sec:esti}. The choice of $N$ is then implied by the type of interpolation  performed after the CDF has been evaluated at the discrete sequence provided by $x_j$.
The number of records $R$ instead, affects directly the variance of the estimator at the value $x_j$, as shown in (\ref{varfirst}).

\subsection{Simulation results}
\label{subs:simulation}
Simulation results using the Monte Carlo method are shown in this subsection both when assuming $s_n$ known and unknown. 
In both cases, the transition levels are assumed to be known. 
If they are unknown, as assumed in subsection 
~\ref{subs:expres}, well grounded 
estimator{\color{black}s} exist that can be used in a preliminary DAQ calibration phase 
\cite{IEEE1241}\cite{Alegria}\cite{CarboneMoschitta}. 

Results of the Monte Carlo simulations 
are shown in Fig.~\ref{gau} and in Fig.~\ref{montegau}.
These are based on $R=1000$ records of $N=151$ 
noisy samples, quantized
assuming an $8$-bit uniform quantizer, 
with a quantization step equal to $\Delta=\frac{2}{2^8}$.
{\color{black} While the assumption on the 
uniform distribution of transition levels 
provides information on the estimator performance 
when applied to nominally ideal ADCs,
(\ref{first}) can be applied to any quantizer, regardless of this assumption. 
In fact, (\ref{first}) does not require or imply any specific sequence of values for 
the transition levels $T_k$.
At the same time, 
experimental results described in Sect.~\ref{experiment}, are obtained in practice, when the regularity hypothesis in the distribution of transition levels is inapplicable.}

\subsubsection{Known AC input sequence} 
\label{subsub:123}
Under the assumption that the input 
sequence was known, we considered a sine wave signal defined as:
\begin{equation}
	s_n=A\sin\left(2\pi \frac{\lambda}{N} n+\phi_0\right), \qquad n=0, \ldots, N-1
	\label{seq}
\end{equation}
where $A=5.37\Delta$ represents the sine wave amplitude, $\lambda=35$ the number of observed periods, $\phi_0=\frac{11}{2}\pi$ the initial record phase and $N=151$ the number of observed samples.
By further assuming zero-mean Gaussian noise with $\sigma=0.25\Delta$ 
the estimated CDF is shown in Fig.~\ref{gau}(a) together with
the reference Gaussian CDF (thin line). The 
good agreement between the two curves sustains the goodness of this approach. 
The estimated CDF is fitted to a Gaussian CDF,
by using a nonlinear iterative least squares estimation procedure.
Accordingly, the mean value was estimated as being equal to  $-1.62\cdot 10^{-6}$ while the standard deviation was estimated as being $0.2515\Delta$.
By using these parameters the estimated noise PDF was plotted in 
Fig.~\ref{montegau}(a) (red solid line) together with the used Gaussian PDF (black solid line). The negligible error between the two curves is plotted in the inset.

\subsubsection{Unknown AC input sequence} 
Simulations were then run to determine how 
the estimator performed when $s_n$ was unknown, apart from $\lambda$.
The same sequence as in (\ref{seq}) was applied to the DAQ under the same conditions listed in \ref{subsub:123}.
Thus, at first, an estimate of $\hat{s}_n$ of $s_n$ 
was obtained by applying the least-squares approach to data quantized by the DAQ.
{\color{black} Accordingly,  
by defining \cite{IEEE1241}
\begin{align}
\setstretch{1.3}
	H=
	\left[
		\begin{array}{ccc}
			0 & 1 & 1 \\
			\sin\left(2\pi \lambda \frac{1}{N}\right) & \cos\left(2\pi \lambda \frac{1}{N}\right) & 1 \\ 
			\sin\left(2\pi \lambda \frac{2}{N}\right) & \cos\left(2\pi \lambda \frac{2}{N}\right) & 1 \\ 
			\vdots & \vdots & \vdots \\
			\sin\left(2\pi \lambda \frac{N-1}{N}\right) & \cos\left(2\pi \lambda \frac{N-1}{N}\right) & 1 \\ 		
		\end{array}
	\right]
\end{align}
for each record $r=0, \ldots, R-1$, we obtain
\begin{align}
	\hat{\theta}_r = (H^T H)^{-1}H^T Y_r
	\quad 
	\hat{\theta}_r = \left[ 
		\begin{array}{c}
			\hat{\theta}_{1r} \\
			\hat{\theta}_{2r} \\
			\hat{\theta}_{3r} 
		 \end{array}
  	\right]
	Y_r = \left[ 
		\begin{array}{c}
			y(0,r) \\
			y(1,r) \\
			\vdots \\
			y(n-1,r) 
		 \end{array}
  	\right]
\end{align}
so that estimates $\hat{A}$ 
of $A$ and $\hat{\phi}_0$ of $\phi_0$ in (\ref{seq}) result as:
\begin{align}
\begin{split}
	& \hat{A} = \sqrt{\overline{\theta}_1^2 +\overline{\theta}_2^2},  \qquad
	 \hat{\phi}_0 = \arctan{\frac{\overline{\theta}_1}{ \overline{\theta}_2}} \\
 & \overline{\theta}_i = \frac{1}{R}\sum_{r=0}^{R-1} \hat{\theta}_{ir}, \quad i=1,2
\end{split}
\end{align}
}
The estimated sequence {\color{black} obtained by substituting $\hat{A}$ and
$\hat{\phi}_0$ in (\ref{seq})} was then treated as if {\color{black} it} were the known input sequence, so that the same CDF estimator could again be applied as in \ref{subsub:123}.


Results are plotted in Fig.~\ref{gau}(b) and (c), from which it can be observed that:
\begin{itemize}
\item even if $s_n$ is unknown, acceptable results are obtained
also because the errors appear to be erratically bouncing around the CDF reference value, so that the nonlinear fit accurately 
recovers the parameters of the original CDF;
\item if a sufficiently large number of records is collected so that the uncertainty due to $\epsilon_y$ can be neglected,
bounds on the estimation errors are indicated by the red lines, evaluated as $\hat{F}_H(\hat{x}_j-\Delta_{\epsilon})$ and
$\hat{F}_H(\hat{x}_j-\Delta_{\epsilon})$, where $\Delta_{\epsilon}$ is calculated numerically in the simulations.
The  aperture of the bounded zone largely depends on the bias introduced by the sinefit: if the noise standard deviation is small compared to $\Delta$, the bias is known to be large \cite{CarboneSchoukens}, as confirmed by the larger width in Fig.~\ref{gau}(c) than in Fig.~\ref{gau}(b). By increasing the noise standard deviation the bias drops and the two bounds tend to collapse, meaning that $\Delta_{\epsilon} \rightarrow 0$;
\item while in this case the estimated CDF was fitted nonlinearly to a Gaussian CDF to show the capability of the estimator to recover its behavior, other parametric or {\color{black} nonparametric} approaches can be taken to interpolate between samples $x_j$ and to then recover the correponding PDF \cite{Sheather}.
\end{itemize}
Once parameters were extracted by using the fitting procedure, PDFs
were calculated and plotted 
in Fig.~\ref{montegau}(b) and (c) using black stars and red lines for the true and estimated PDF, respectively.
With respect to Fig.~\ref{montegau}(a), the input sequence $s_n$ was first estimated by determining the values of the amplitude, offset and initial record phase of the input sinewave, using a sinefit procedure \cite{IEEE1241}. Even in this case the estimation error can be considered acceptable for most applications.

\begin{figure}
\begin{center}
\includegraphics[scale=0.45]{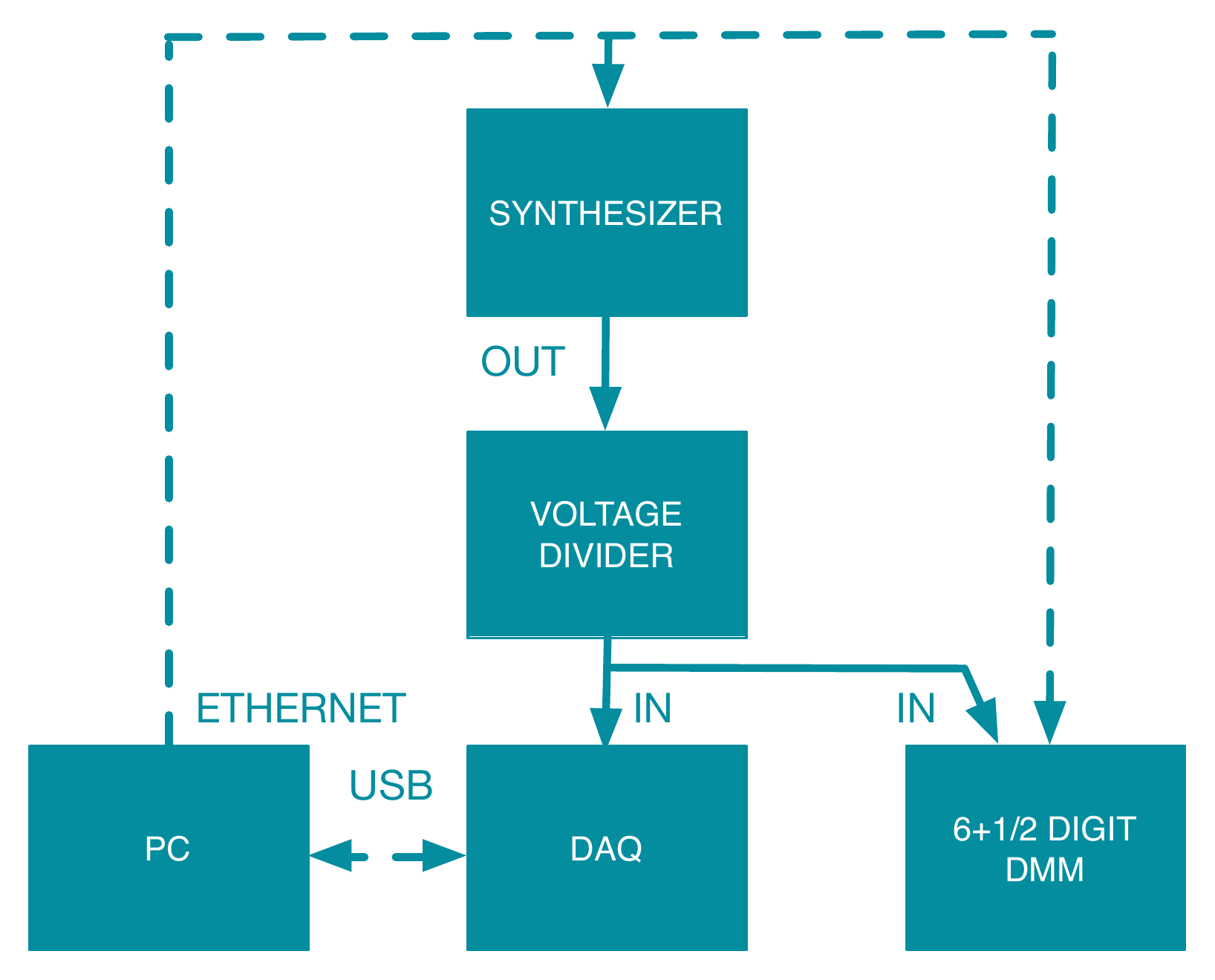}
\caption{The measurement set-up used to obtain experimental data.
The synthesizer sources a sine wave via a voltage divider to the $12$-bit DAQ, whose amplitude is measured by the digital multimeter in AC mode. A personal computer uses Ethernet and the Universal Serial Bus (USB) to control the measurement chain.
\label{expres}}
\end{center}
\end{figure}

\subsection{Experimental results \label{experiment}}
\label{subs:expres}
\begin{figure}
\includegraphics[scale=0.45]{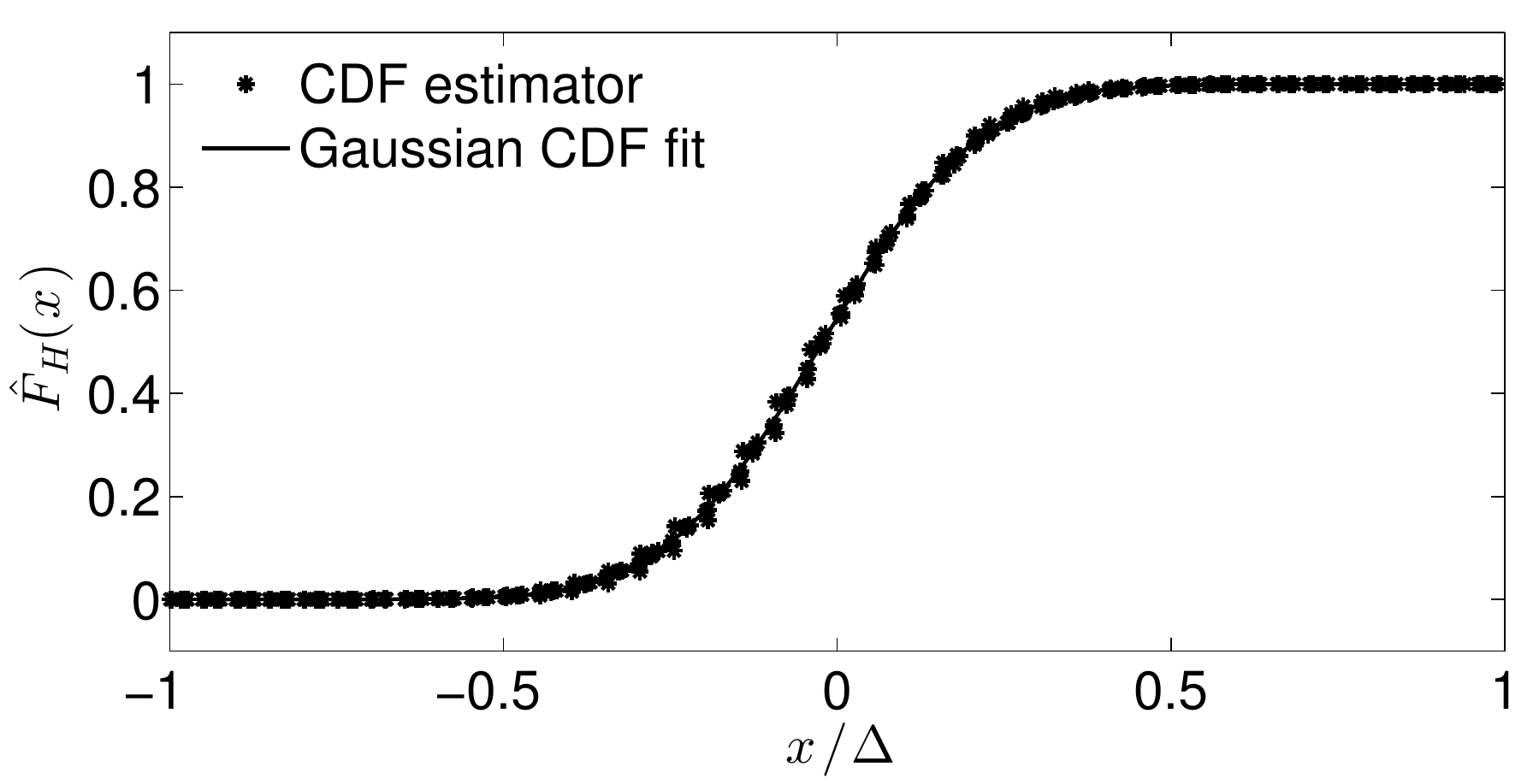}
\caption{Experimental results obtained using the set-up shown in Fig.~\ref{expres}: estimated noise CDF (stars) and fitted Gaussian CDF (solid line). \label{expcdf}}
\end{figure}

Measurement data were collected to validate 
the proposed estimator using a swept DC input sequence.
A Keysight U2331 $12$-bit data acquisition system was connected to a personal computer, while a $6\nicefrac{1}{2}$ digit multimeter was used to provide a reference value of the DAQ input signal, as shown in Fig.~\ref{expres}. A $14$-bit resolution synthesizer was used to source the 
DAQ via a voltage divider having an approximate ratio of $10:1$. The exact ratio value was not needed as the DMM was used to provide the measured values.
The procedure was applied in two steps:
\begin{itemize}
\item at first the DAQ was calibrated, that is values of the transition levels were measured. To this aim the PC was programmed to implement 
the {\em servoloop} method \cite{IEEE1241}.
Accordingly, for any given DAQ code, the synthesizer was programmed to provide that DC value causing the DAQ 
output to be half of the time above and half below the given code. Once this condition was achieved, the DMM was used to measure the provided DC value. By iterating this technique over several DAQ codes a set of transition level values was collected. 
The values of transition levels were then fitted to a straight line. 
It was found that the
absolute maximum error between the transition level position and the linear fit, over the measured voltage span, was bounded by $\nicefrac{\Delta}{4}$;
\item having information on the position of the transition levels, the synthesizer was used to scan the DAQ input value over the interval $[-4\Delta, 4\Delta]$, with $\Delta=20/2^{12}=0.004882 \ldots$ V,
in steps of about $2.45\cdot10^{-4}$ V. Each time the provided value was also 
measured by the DMM.
For each input value, $2.5\cdot 10^5$ samples were collected by the DAQ and processed according to the estimator defined in (\ref{first}). 
Results are shown in Fig.~\ref{expcdf} using dots.
\end{itemize}
Given the appearance of the estimated CDF, data were fitted to a Gaussian CDF, also plotted in Fig.~\ref{expcdf} using a solid line. The maximum absolute error between the fitted curve and the data, was observed to be bounded by $0.03$. The corresponding estimated mean value and standard deviations were found to be equal to $-0.0214\Delta$ and
$0.1867\Delta$, respectively.


\section{Conclusion}
In this paper we described a simple estimator for measuring the noise CDF and PDF at the input
of a quantizer inside an ADC or a data acquisition board.
The approach taken is based on considering the entire amount of information returned by the
used ADC and not just that carried by the output codes. This implies that processing takes into account
the values of the transition levels in the used ADC to improve estimates over the usage of a simple 
empirical distribution function based on the ADC output codes. As simulation and 
experimental results show, the estimator proposed in this paper 
provides good results both when assuming uniform and non-uniform ADC transition levels and when considering DC and AC input sequences.

\section*{Acknowledgement}
This work was supported in part by the Fund for Scientific Research (FWO-Vlaanderen), by the Flemish Government (Methusalem), the Belgian Government through the Inter university Poles of Attraction (IAP VII) Program, and by the ERC advanced grant SNLSID, under contract 320378.

\balance


\begin{thebibliography}{12}
\bibitem{IEEE1241} IEEE, {\em Standard for Terminology and Test Methods for Analog-to-Digital Converters}, IEEE Std. 1241, Aug. 2009.
\bibitem{AD9265}Analog Devices, AD9265 Data Sheet Rev C, 08/2013.
\bibitem{Wagdy}M.F. Wagdy, 
``Effect of Additive Dither on the Resolution of ADC's with Single-Bit or Multibit Errors,''
{\em IEEE Trans. Instr. Meas.}, vol. 45, pp. 610-615, April 1996.
\bibitem{CarboneNarduzziPetri}P. Carbone, C. Narduzzi, D. Petri, ``Dither signal 
effects on the resolution of nonlinear quantizers,''
{\em IEEE Trans. Instr. Meas.}, vol. 43, no. 2, pp. 139-145, Apr. 1994.
\bibitem{Lee} Leee Byung-Uk, ``Modelling quantisation error from quantised Laplacian distributions,'' 
{\em Electronics Letters}, vol. 36, no. 15, pp. 1270-1271, July 2000.
\bibitem{CarboneMoschittaPDF}A. Moschitta, P. Carbone, ``Noise Parameter Estimation From Quantized Data,'' 
{\em IEEE Trans. Instr. Meas.}, vol. 56, no. 3, pp. 736-742, June 2007.
\bibitem{Chiuso} A. Chiuso, ``A note on estimation using quantized data,'' Proc. of the 17th World Congress
The International Federation of Automatic Control Seoul, Korea, July 6-11, 2008.
\bibitem{WangYinZhangZhao}L. Y. Wang, G. G. Yin, J. Zhang, Y. Zhao, {\em  System Identification with Quantized Observations,} Springer Science, 2010.
\bibitem{104} L.~Y.~Wang and G.~Yin, ``Asymptotically efficient parameter estimation using quantized output observations,'' {\em Automatica}, Vol. 43, 2007, pp. 1178-1191.
\bibitem{Qli}Q. Lia, J. Racineb, ``Nonparametric estimation of distributions with categorical and continuous data,''
{\em Journal of Multivariate Analysis}, vol. 86, pp. 266-292, 2002.
{\color{black}
\bibitem{Agresti}A. Agresti,
{\em Analysis of Ordinal Categorical Data, 2nd Edition}, Wiley, 2010, ISBN: 978-0-470-08289-8.
\bibitem{Kurt}K.~Barb\'e, L.~G.~Fuentes, L.~Barford, L.~Lauwers, ``A Guaranteed Blind and Automatic Probability Density Estimation of Raw Measurements,''  {\em IEEE Trans. Instrum. Meas.}, vol. 63, no. 9, pp. 2120-2128, Sept. 2014.
\bibitem{Masry} E. Masry,  ``Probability density estimation from sampled data,''  {\em IEEE Trans. Inform. Theory}, vol.29, no.5, pp.696-709, Sept. 1983.
\bibitem{Parzen}E.~Parzen, 
``On the estimation of a probability density function and the mode,''
{\em Annals of Math. Stats.}, vol. 33, pp. 1065-1076, 1962.
}
\bibitem{Alegria}F.~Correa~Alegria,
``Bias of amplitude estimation using three-parameter sine fitting in the presence of additive noise,''
 {\em IMEKO Measurement}, 2 (2009), pp. 748-756.
 \bibitem{HandelLSE}P.~Handel,
``Amplitude estimation using IEEE-STD-1057 three-parameter sine wave fit: Statistical distribution, bias and variance,'' {\em IMEKO Measurement}, 43 (2010), pp. 766-770.
\bibitem{CarboneSchoukens}P. Carbone, J. Schoukens, ``A Rigorous Analysis of Least Squares Sine Fitting
Using Quantized Data: The Random Phase Case,''
{\em IEEE Trans. Instr. Meas.}, vol. 63, pp. 512-529, March 2014.
\bibitem{Kollar1}
L. Balogh, I. Koll\'ar, L. Michaeli, J. \v{S}aliga, J. Lipt\'ak,
``Full information from measured ADC test data using maximum likelihood estimation,'' {\em Measurement}, vol. 45, pp. 164-169, 2012.
\bibitem{Kollar2}
J. \v{S}aliga, I. Koll\'ar, L. Michaeli, J. Bu\v{s}a, J. Lipt\'ak, T. Virosztek, 
``A comparison of least squares and maximum likelihood methods using sine fitting in ADC testing,'' {\em Measurement}, vol. 
46, pp. 4362-4368, 2013.
\bibitem{Sheather}Simon J. Sheather, ``Density Estimation,'' {\em Statistical Science}, no. 4, vol. 19, pp. 588-597, 2004. 
\bibitem{CarboneMoschitta} A. Moschitta, P. Carbone, 
``Cram\'er-Rao lower bound for parametric estimation of quantized sinewaves,'' 
{\em IEEE Trans. Instr. Meas.} June 2007, vol. 56, no. 3, pp. 975-982.
%
\end{thebibliography}
\end{document}